\input harvmac
\overfullrule=0pt

\Title{ \vbox{\baselineskip12pt
\hbox{hep-th/0303007}}}
%\hbox{IASSNS-HEP-99/66}
%\hbox{CAL}}}
{\vbox{\centerline{Graviton-Scalar Interaction}
\bigskip
\centerline{in the PP-Wave Background}}}
\smallskip
\centerline{K. Bobkov
\foot{bobkov@physics.unc.edu}
}
\smallskip
\centerline{\it Department of Physics}
\centerline{\it
University of North Carolina, Chapel Hill, NC 27599-3255}
\bigskip
\smallskip

\noindent
We compute the graviton two scalar off-shell interaction vertex at tree level in 
Type IIB superstring theory on the pp-wave background using the light-cone
string field theory formalism. We then show that the tree level vertex 
vanishes when all particles are on-shell and conservation of $p_+$ and 
$p_ -$ are imposed. We reinforce our claim by calculating the same 
vertex starting from the corresponding SUGRA action expanded around the 
pp-wave background in the light-cone gauge.
\Date{}
%references
\nref\mald{J.M. Maldacena,''The Large N Limit of Superconformal Field Theories
and Supergravity'', Adv. Theor. Math. Phys. {\bf 2} (1998) 231 [Int. J. Theor. Phys.
{\bf} (1999) 1113]; arXiv:hep-th/9711200.}
\
\nref\gkp{S.S. Gubser, I.R. Klebanov, A.M. Polyakov, ``Gauge Theory Correlators from 
Non-Critical String Theory'', Phys. Lett. {\bf B428} (1998) 105; arXiv:hep-th/9802109}
\
\nref\wit{E. Witten, ``Anti-de Sitter Space and Holography'', Adv. Theor. Math. Phys.
{\bf 2} (1998) 253; arXiv:hep-th/9802150.}
\
\nref\bfhpone{M.Blau, J. Figueroa-O`Farill, C. Hull, G. Papadopoulos, ``A New
Maximally Supersymmetric Background of IIB Superstring Theory'', JHEP {\bf 0201}
(2002) 047; arXiv:hep-th/0110242.}
\
\nref\bfhptwo{M.Blau, J. Figueroa-O`Farill, C. Hull, G. Papadopoulos, ``Penrose
Limits and Maximal Supersymmetry'', Class. Quant. Grav. {\bf 19}
(2002) L87; arXiv:hep-th/0110242.}
\
\nref\m{R.R. Metsaev, ``Type IIB Green-Schwarz Superstring in Plane Wave
Ramond-Ramond Background'', Nucl. Phys. {\bf B625} (2002) 70; arXiv:hep-th/0112044.}
\
\nref\mt{R.R. Metsaev and A.A. Tseytlin, ``Exactly Solvable Model of 
Superstring in Plane Wave Ramond-Ramond Background'', 
Phys.Rev.{\bf D65} (2002) 126004; arXiv:hep-th/0202109.}
\
\nref\bmn{D. Berenstein, J.M. Maldacena, H. Nastase, ``Strings in Flat Space
and PP-Waves from N=4 Super Yang Mills'', JHEP {\bf 0204} (2002) 013; 
arXiv:hep-th/0202021.}
\
\nref\kpss{C. Kristjansen, J. Plefka, G.W. Semenoff, M. Staudacher, ``A New Double 
Scaling Limit of N=4 Super Yang-Mills Theory and PP Wave Strings'', Nucl. Phys. {\bf B643} 
(2002) 3; arXiv:hep-th/0205033}
\
\nref\rmg{D.J. Gross, A. Mikhailov, R. Roiban, ``Operators with Large R Charge in N=4 
Yang-Mills Theory'', Annals Phys. {\bf 301} (2002) 31; arXiv:hep-th/0205066.}
\
\nref\sanz{A.Santambrogio, D. Zanon, ``Exact Anomalous Dimensions of N=4 Yang-Mills Operators
with Large R Charge'', Phys. Lett. {\bf B545} (2002) 425; arXiv:hep-th/0206079.}
\
\nref\cfhmmp{N.R. Constable, D.Z. Freedman, M. Headrick, S. Minwala, L. Motl, A. Postnikov,
W. Skiba, ``PP-Wave String Interactions from Perturbative Yang-Mills'', JHEP {\bf 0207}
(2002) 017; arXiv:hep-th/0205089.}
\
\nref\gs{M.B. Green and J.H. Schwarz, ``Extended Supergravity In 
Ten Dimensions'', Phys. Lett. {\bf B122} (1983) 143.}
\
\nref\gsone{M.B. Green and J.H. Schwarz, ``Superstring Interactions'', 
Nucl. Phys. {\bf B218} (1983) 43.}
\
\nref\gsb{M.B. Green, J.H. Schwarz, L.Brink, ``Superfield Theory Of Type II
Superstrings'', Nucl. Phys. {\bf B219} (1983) 437.}
\
\nref\svone{M. Spradlin and A. Volovich, ``Superstring Interactions
in PP wave Background'',
Phys.Rev.{\bf D66}(2002)086004; arXiv:hep-th/0204146.}
\
\nref\svtwo{M. Spradlin and A. Volovich, ``Superstring Interactions
in PP wave Background II'',JHEP {\bf 0301}(2003)036; arXiv:hep-th/0206073.}
\
\nref\pankstef{A. Pankiewicz and B. Stefanski, Jr. ,``PP Wave Light Cone
Superstring Field Theory''; arXiv:hep-th/0210246.}
\
\nref\jhs{J.H. Schwarz, ``Comments on Superstring Interactions in a Plane Wave
Background'', JHEP {\bf 0209} (2002) 058; arXiv:hep-th/0208179.}
\
\nref\pank{A. Pankiewicz, ``More Comments on Superstring Interactions in PP-Wave
Background'', JHEP {\bf 0209} (2002) 056; arXiv:hep-th/0208209.}
\
\nref\svs{Y. He, J.H. Schwarz, M. Spradlin, A. Volovich,
``Explicit Formulas for Neumann Coefficients in the Plane Wave 
Geometry''; arXiv:hep-th/0211198.}
\
\nref\huang{M.x. Huang, ``Three Point Functions of N=4 Super Yang Mills from
Light Cone String Field Theory in PP-Wave'',Phys. Lett. {\bf B542}(2002)255;
arXiv:hep-th/0205311.}
\
\nref\ckt{C.S. Chu, V.V. Khoze, G. Travaglini, ``Three-Point Functions in
N=4 Yang Mills Theory and PP-Waves'', JHEP {\bf 0206} (2002) 011; arXiv:hep-th/0206005.}
\
\nref\kklp{Young-jai Kiem, Yoon-bai Kim, Sang-min Lee, Jae-mo Park , 
``PP Wave / Yang-Mills Correspondence: an Explicit Check'', 
Nucl.Phys.{\bf B642}(2002) 389; arXiv:hep-th/0211073.}
\
\nref\lmpone{P.L. Lee, S. Moriyama and J. Park, ``Cubic Interactions in PP-Wave
Light Cone String Field Theory'',Phys. Rev.{\bf D66} {2002} 085021; arXiv:hep-th/0206065.}
\
\nref\lmptwo{P.L. Lee, S. Moriyama and J. Park, ``A Note On Cubic Interactions 
in PP-Wave Light Cone String Field Theory'', arXiv:hep-th/0209011.}
\
\nref\last{Young-jai Kiem, Yoon-bai Kim and Jaemo Park,``Chiral Primary
Cubic Interactions From PP-Wave Supergravity'', JHEP {\bf 0301} (2003) 026;
arXiv:hep-th/0211217.}
\
\nref\dbsj{Dong-su Bak, Mohammad M. Sheikh-Jabbari, ``Strong Evidence in
Favor of the Existence of S Matrix for Strings in Plane Waves'', 
JHEP {\bf 0302} {2003} 019; arXiv:hep-th/0211073.}
\
\nref\gr{I.S. Gradshtein, I.M. Ryzhik, ``Table of integrals, series,
and products'', Academic Press, San Diego, 2000.}
\
%paper
\newsec{Introduction}
In \refs{\mald} it was conjectured that there exists a duality between the
Type IIB string theory on $AdS_5\times S^5$ and ${\cal N}=4$ super Yang Mills theory
on the boundary of $AdS_5$. A concrete recipe for verifying this correspondence 
in the large $\lambda\equiv g^2_{YM}\,N\rightarrow \infty$ limit when 
$\alpha^{\prime}\rightarrow 0$ was given in \refs{\gkp-\wit}. There have been 
many checks of this duality in the supergravity ($\alpha^{\prime}\rightarrow 0$) 
limit but the verification of the correspondence in the full string theory 
remains elusive. In a more recent development, the authors of 
\refs{\bfhpone-\bfhptwo} showed that in the Penrose
limit the $AdS_5\times S^5$ background turns into the pp-wave solution of
Type IIB supergravity preserving all 32 supercharges
\eqn\metric{ds^2=-4dx^{+}dx^{-}-\mu^2x_Ix_I\left(dx^{+}\right)^2
+dx_Idx_I\,,\,\,\,\,\,\,F_{+1234}=F_{+5678}=2\mu\,,}
where $I=1,...,8$.
Unlike the case of $AdS_5\times S^5$, where we do not even know the free string 
spectrum, string theory on the pp-wave background can be solved \refs{\m-\mt} 
in the light-cone gauge, despite the presence of a non-zero Ramond-Ramond flux. 
Partly motivated by the conjecture in \refs{\mald}, Berenstein, Maldacena and 
Nastase \refs{\bmn} have argued that a particular sector of ${\cal N}=4$ super 
Yang Mills theory containing operators with large R-charge J is dual to Type IIB 
string theory on the pp-wave background with Ramond-Ramond flux. The fact that 
string theory on the pp-wave background has been exactly solved has opened an 
exciting possibility to check the proposed correspondence beyond the supergravity 
limit. In fact, the authors of \refs{\bmn} succeeded in reproducing the tree level 
string spectrum on the pp-wave \refs{\mt} from the perturbative super Yang Mills 
theory as the first test of a full string theory/CFT duality. The anomalous 
dimensions of the BMN operators arising from the Yang-Mills perturbation theory 
were studied in \refs{\kpss-\cfhmmp} to check the correspondence between strings
on the plane wave background and the Yang-Mills theory at the level of perturbative 
expansions. In a separate development, Spradlin and Volovich generalized the 
formalism of the light-cone string field theory in Minkowski space \refs{\gs-\gsb} 
to the plane-wave geometry \refs{\svone-\svtwo}. The subject of the pp-wave
light-cone string field theory pioneered by SV in \refs{\svone-\svtwo} was studied 
further in \refs{\pankstef-\pank}.
The factorization theorem for the Neumann coefficients was discussed in 
\refs{\jhs-\pank} and explicit formulas for the Neumann coefficients were 
derived in \refs{\svs}. By employing the formalism of  \refs{\svone-\svtwo}, 
certain three-string amplitudes  were found to be in agreement with the corresponding
three-point functions of the BMN operators with large R charge \refs{\svtwo}, 
\refs{\huang-\lmptwo}. A cubic light-cone interaction Hamiltonian for the chiral 
primary system  from the Type IIB pp-wave supergravity  was constructed 
in \refs{\kklp}, \refs{\last}.
Another interesting question is the issue of existence of the 
$S$-matrix interpretation for field theories on the plane wave backgrounds.
The authors of \refs{\dbsj} have demonstrated that, at least at the tree level, 
the field theory of scalars and scalars coupled to a gauge field do have
an $S$-matrix formulation.

The motivation behind this paper is brought about by the fact that, unlike in
the flat Minkowski space, the stringy modes do not decouple from the supergravity
modes in the pp-wave background \refs{\huang-\lmpone}. Therefore, there 
is a possibility that the $\alpha^{\prime}$ corrections could potentially 
contribute to the three-point interactions at tree level. In section 2 we 
compute the string tree level three-point amplitude for a graviton and two scalars,
i.e. we use the formalism of \refs{\svone-\svtwo} to compute a 
matrix element of the cubic interaction Hamiltonian for the states represented 
by the graviton and a combination of the dilaton and axion fields of the 
Type IIB supergravity on the pp-wave background. We discover that the off-shell
amplitude contains $\alpha^{\prime}$ corrections encoded in the Neumann
coefficients of the zero modes of the full string theory vertex. We show that
when we impose conservation of $p_{+}$ and $p_{-}$, this amplitude vanishes
on-shell. In section 3, we compute the graviton dilaton-axion cubic vertex
starting from the Type IIB supergravity action expanded around the pp-wave 
background in the light-cone gauge. We show that it vanishes on-shell and
thus verify in the $\alpha^{\prime}\rightarrow 0$ limit our string theory
computation.
\vskip 20pt
\newsec{Graviton axion-dilaton interaction from light cone string field theory}
\subsec{Some Key Results of the Light-Cone String Field Theory in PP-Wave}
Following the light-cone string field theory formalism of \refs{\gs-\gsb} 
developed in Minkowski space, Spradlin and Volovich successfully generalized
it to strings propagating in the plane wave background \refs{\svone-\svtwo}. 
In particular, they constucted a cubic interaction Hamiltonian that can
be expressed as
\eqn\ffive{|H_3\rangle=\hat h_3|V\rangle\,,}
where $|V\rangle$ is a three-string vertex satisfying the kinematic
constraints given by equations (4.2)-(4.5)
of  \refs{\svone} and the prefactor 
$\hat h_3$ must be inserted at the interaction point in order to 
preserve supersymmetry. More explicitly
\eqn\fesix{|V\rangle=E_aE_b|0\rangle\,,}
where the bosonic and fermionic zero mode parts of the vertex are
\eqn\feseven{E^0_a={\rm exp}\left[{1\over 2}\sum_{r,s=1}^3\sum_{I=1}^8
a^{\dagger I}_r\overline N^{rs}a^{\dagger I}_s\right]\,,}
\eqn\feight{E^0_b={1\over 8!}\epsilon_{a_{1}...a_{8}}\hat\lambda^{a_{1}}...
\hat\lambda^{a_{8}},\,\,\,{\rm where}\,\, \hat\lambda^a=\hat\lambda_1^a+
\hat\lambda_2^a+\hat\lambda_3^a,}
$\hat\lambda^a$'s are zero modes of the fermionic conjugate 
momenta of the string. They are complex positive chirality 
$SO(8)$ spinors with $a=1,...,8$. The strings are labeled by 
$r,\,s=1,\,2,\,3$.
We are only going to be concerned with the states corresponding to the
Type IIB supergravity multiplet \mt. However, as was pointed out to us 
by M. Spradlin and was discussed in \refs{\huang-\lmpone}, the 
supergravity modes do not decouple from the string modes in the 
pp-wave background. Decoupling only takes place in the flat space
limit when $\mu\alpha^{\prime}p^{+}\rightarrow 0$. We will therefore be 
using the Neumann coefficients for the zero modes of the full string 
theory vertex derived explicitly in \refs{\svs}, instead of the 
supergravity vertex given in \refs{\svone}. The zero mode Neumann coefficients are
\eqn\fenine{\eqalign{&\overline N^{rs}=(1+\mu\alpha k)\epsilon^{rt}\epsilon^{su}
\sqrt{\alpha_t\alpha_u\over\alpha^2_3},\,\,\,r,\,s,\,t,\,u\,\in \{1,2\},\cr
&\overline N^{3r}=\overline N^{r3}=-\sqrt{-{\alpha_r\over\alpha_3}},\,\,\,\,\,r\in 
\{1,2\},\cr &\overline N^{33}=0,}}
where $\alpha=\alpha_1\alpha_2\alpha_3$ and $\alpha_r\equiv 2p^{+}_r$ and where
by $p^{+}$ we really mean $-p_{-}$. This point is important because in the 
pp-wave metric we have a non-zero $\tilde g^{++}$ component and therefore
strictly speaking $p^{+}$ and  $-p_{-}$ are different.
From eq.(4.18) of \refs{\svone} we have the following prefactor for the 
zero modes
\eqn\feten{\hat h_3={\rm P}^I{\rm P}^Jv_{IJ}\left(\Lambda\right),}
where
\eqn\gone{v_{IJ}\left(\Lambda\right)=\delta^{IJ}+{1\over 6\alpha^2}t^{IJ}_{abcd}
\Lambda^a\Lambda^b\Lambda^c\Lambda^d+{16\over 8!\alpha^4}\delta^{IJ}
\epsilon_{abcdefgh}\Lambda^a\Lambda^b\Lambda^c\Lambda^d\Lambda^e
\Lambda^f\Lambda^g\Lambda^h,}
\eqn\gtwo{\Lambda^a=\alpha_1\hat\lambda_2^a-\alpha_2\hat\lambda_1^a
=\alpha_3\hat\lambda_1^a-\alpha_1\hat\lambda_3^a=
\alpha_2\hat\lambda_3^a-\alpha_3\hat\lambda_2^a,}
\eqn\gthree{{\rm P}^I=\alpha_1\hat p_2^I-\alpha_2\hat p_1^I=
\alpha_3\hat p_1^I-\alpha_1\hat p_3^I=\alpha_2\hat p_3^I-\alpha_3\hat p_2^I\,,}
and
\eqn\xone{\hat p^I=\sqrt{|\alpha|\mu}\left(a^I+a^{\dagger I}\right)\,.}
The self-dual tensor $t^{IJ}_{abcd}$ was defined in \refs{\gs} in 
terms of $SO(8)$ gamma matrices as 
\eqn\gfour{t^{IJ}_{abcd}=\gamma^{IK}_{[ab}\gamma^{JK}_{cd]}}
and satisfies various identities given in Appendix A of \refs{\gsb}.
Appendix C of this paper contains some extra identities for tensor
$t^{IJ}_{abcd}$ that could be useful in future computations.
Using the constraint $\left(\hat p_1^I+\hat p_2^I+\hat p_3^I\right)|V\rangle=0$
together with $\alpha_1+\alpha_2+\alpha_3=0$ it is easy to show that 
\eqn\fften{{\rm P}^I{\rm P}^J|V\rangle=
-\alpha_1\alpha_2\alpha_3\left[{{1\over\alpha_1}\hat p^I_1\,\hat p^J_1+
{1\over\alpha_2}\hat p^I_2\,\hat p^J_2+{1\over\alpha_3}\hat p^I_3\,\hat p^J_3}
\right]|V\rangle\,.}
Following \svone\ will assume that $\alpha_1$ and $\alpha_2$ are positive
with $\alpha_1+\alpha_2+\alpha_3=0$. 
We can further substitute \xone\ for $\hat p^I_r$ to obtain from \fften\
the following
\eqn\ffeleven{\eqalign{{\rm P}^I{\rm P}^J|V\rangle=
-\mu\alpha_1\alpha_2\alpha_3&\biggl[\left(a_1^I+a_1^{\dagger I}\right)
\left(a_1^J+a_1^{\dagger J}\right)+\left(a_2^I+a_2^{\dagger I}\right)
\left(a_2^J+a_2^{\dagger J}\right)\cr
&-\left(a_3^I+a_3^{\dagger I}\right)
\left(a_3^J+a_3^{\dagger J}\right)\biggr]|V\rangle\,.}}
For $I=J$ \ffeleven\ can be written \refs{\lmpone} as
\eqn\fftwelve{{\rm P}^I{\rm P}^I|V\rangle=
-\mu\alpha_1\alpha_2\alpha_3\left[a_1^{\dagger I}a_1^I
+a_2^{\dagger I}a_2^I-a_3^{\dagger I}a_3^I\right]|V\rangle\,.}
The light-cone Hamiltonian for bosonic zero modes is given by
\eqn\rone{H_r=\mu\,\sum_{I=1}^8a_r^{\dagger I}a_r^I+\mu\,E^r_0\,,}
and the light-cone energy is
\eqn\rtwo{p^r_{+}=\mu\,\sum_{I=1}^8\,n^r_I+\mu\,E^r_0\,.}
\subsec{Graviton Dilaton-Axion Vertex}
Here we will calculate a three string amplitude using the formalism
of \refs{\svone} for a particular choice of states from the Type IIB 
supergravity multiplet. The superfield expansion for Type IIB supergravity 
in light-cone gauge
originally given by equation (1) of \refs{\gs} is
\eqn\fone{\eqalign{\Phi(x,\theta)&=\sum_{N=0}^4{1\over\left(2N\right)!}
\left(\partial^{+}\right)^{N-2}A_{a_{1}a_{2}...a_{2N}}\theta^{a_{1}}
\theta^{a_{2}}...\theta^{a_{2N}}\cr
&+\sum_{N=0}^3{1\over\left(2N+1\right)!}\left(\partial^{+}
\right)^{N-2}\psi_{a_{1}a_{2}...a_{2N+1}}\theta^{a_{1}}\theta^{a_{2}}...
\theta^{a_{2N+1}}\,.}}
We are going to be interested in the bosonic terms corresponding
to $N=0,2,\,{\rm and}\,\,4$ of the first sum that contain the dilaton, 
axion and graviton.
\eqn\ftwo{\eqalign{\Phi(x,\theta)&={1\over\left(\partial^{+}\right)^2}
A^{*}(x)+{1\over\ 4!}A_{abcd}(x)\theta^a\theta^b\theta^c\theta^d\cr
&+{1\over\ 8!}\left(\partial^{+}\right)^2A(x)\epsilon_{abcdefgh}
\theta^a\theta^b\theta^c\theta^d\theta^e\theta^f\theta^g\theta^h+...\,,}}
where we set $A_{abcdefgh}(x)=A(x)\epsilon_{abcdefgh}$.
The fields of interest are identified \refs{\gsb} as 
\eqn\fthree{\eqalign{&\tau(x)=\chi(x)+i e^{-\phi(x)}=A(x)\,,\cr
&\bar\tau(x)=\chi(x)-i e^{-\phi(x)}=A^{*}(x)\,,\cr
&h^{IJ}(x)={1\over2}\,t^{IJ}_{abcd}\,A^{abcd}(x)\,,}}
where $\chi$ is the RR scalar (axion), $h^{IJ}$ is symmetric
and traceless (graviton), and $\phi$ is the trace (dilaton).
As prescribed by \refs{\svone} it is necessary to transform the superfield
\ftwo\ to the occupation number basis $\{k_I\}$ for the transverse 
directions $x^{I},\,\left(I=1,...,8\right)$ and to momentum space 
for $x^{-}$ coordinate
\eqn\ffour{\eqalign{\Phi\left(x^{+},\alpha,\theta;\{k_{I}\}\right)&={4\over\alpha^2}
\bar\tau\left(x^{+},\alpha;\{k_{I}\}\right)+{1\over\ 4!}A_{abcd}\left(x^{+},\alpha;
\{k_{I}\}\right)\theta^a\theta^b\theta^c\theta^d\cr
&+{\alpha^2\over 4}\tau\left(x^{+},\alpha;\{k_{I}\}\right){1\over 8!}
\epsilon_{abcdefgh}\theta^a\theta^b\theta^c\theta^d\theta^e\theta^f
\theta^g\theta^h+...\,\,,}}
where we used \fthree\ to replace $A$ and $A^{*}$ with $\tau$ and $\bar\tau$.
The expression that we are about to evaluate has the form
\eqn\ffive{\langle\Phi(1)|\langle\Phi(2)|\langle\Phi(3)|H_3\rangle\,,}
where we are only going to be interested in the terms proportional to 
$A_{abcd}\tau\bar\tau$ that contain the graviton coupled to the 
dilaton-axion pair. We will first deal with the fermionic zero modes and
use in our calculation the following conditions on $\hat\theta^a$ and
its conjugate momentum $\hat\lambda^a$
\eqn\fsix{\eqalign{&\hat\theta^a|0\rangle=0,\cr
&\langle0|\hat\lambda^a=0,\cr
&\{\hat\theta^a,\hat\lambda^b\}=\delta^{ab}.}}
By counting the number of $\hat\lambda$'s on the right hand side
to saturate the number of $\hat\theta$'s on the left hand side,
we see that only the second term in \gone\ will contribute to the
$h\tau\bar\tau$ interaction. Suppressing the $x^{+}$ dependence, we
have the following expression for the graviton scalar vertex
\eqn\fseven{\eqalign{A_{h\tau\bar\tau}(1,2,3)&=\langle\{k^1_I\}|
\hat\theta_1^{a_1}\hat\theta_1^{a_2}\hat\theta_1^{a_3}\hat\theta_1^{a_4}
{1\over 4!}A_{a_1a_2a_3a_4}\left(\alpha_1,\{k^1_I\}\right)\,\langle\{k^2_I\}|
{4\over\alpha_2^2}\bar\tau\left(\alpha_2,\{k^2_I\}\right)\cr
&\times\langle\{k^3_I\}|
\hat\theta_3^{b_1}\hat\theta_3^{b_2}\hat\theta_3^{b_3}\hat\theta_3^{b_4}
\hat\theta_3^{b_5}\hat\theta_3^{b_6}\hat\theta_3^{b_7}\hat\theta_3^{b_8}
{1\over 8!}\epsilon_{b_1b_2b_3b_4b_5b_6b_7b_8}{\alpha_3^2\over 4}\tau
\left(\alpha_3,\{k^3_I\}\right)\cr
&\times{\alpha_2^4\over {6(\alpha_1\alpha_2\alpha_3)^2}}t^{JK}_{c_1c_2c_3c_4}
\hat\lambda^{c_1}_1\hat\lambda^{c_2}_1\hat\lambda^{c_3}_1\hat\lambda^{c_4}_1\cr
&\times{1\over 8!}\epsilon_{d_1d_2d_3d_4d_5d_6d_7d_8}
\hat\lambda_3^{d_1}\hat\lambda_3^{d_2}\hat\lambda_3^{d_3}\hat\lambda_3^{d_4}
\hat\lambda_3^{d_5}\hat\lambda_3^{d_6}\hat\lambda_3^{d_7}\hat\lambda_3^{d_8}
{\rm P}^J{\rm P}^KE^{0}_a|0\rangle+c.c.\,\,,}}
where the occupation number states are defined as
\eqn\feight{\eqalign{&|\{k^r_I\}\rangle=\prod_{I=1}^8(-i)^{k^r_I}
{\left(a^{\dagger I}_r\right)^{k^r_I}\over\sqrt{k^r_I!}}|0\rangle\,,\cr
&\langle\{k^r_I\}|=\langle 0|\prod_{I=1}^8(i)^{k^r_I}
{\left(a^I_r\right)^{k^r_I}\over\sqrt{k^r_I!}}\,,\cr
&{\rm \,\,with\,\,}\left[a^I_r,a^{\dagger J}_s\right]=
\delta^{IJ}\delta_{rs}\,.}}
Definitions \feight\ are based on the definitions of $a^I_r$ and $a^{\dagger I}_r$
given in \refs{\svone}. Following the standard procedure to bring all 
the $\hat\lambda^a$'s to the left and all the $\hat\theta^a$'s to
the right and using \fsix, we have from \fseven
\eqn\fnine{\eqalign{A_{h\tau\bar\tau}(1,2,3)&={1\over 3\alpha_1^2}
\langle\{k^1_I\}|\langle\{k^2_I\}|\langle\{k^3_I\}|{\rm P}^J{\rm P}^K
E^{0}_a|0\rangle\cr
&\times{1\over 2}t^{JK}_{c_1c_2c_3c_4}A^{c_1c_2c_3c_4}\left(\alpha_1,\{k^1_I\}\right)
\bar\tau\left(\alpha_2,\{k^2_I\}\right)\tau\left(\alpha_3,\{k^3_I\}\right)+c.c.\,\,\,.}}
We can now identify the graviton in \fnine\ using \fthree\ and use the explicit
representation for the occupation number states given by \feight\ to obtain from
\fnine\
\eqn\ften{\eqalign{A_{h\tau\bar\tau}(1,2,3)&={1\over 3\alpha_1^2}
\prod_{I=1}^8{i^{k^1_I+k^2_I+k^3_I}\over\sqrt{k^1_I!k^2_I!k^3_I!}}
\langle 0|\left(a^I_1\right)^{k^1_I}\left(a^I_2\right)^{k^2_I}
\left(a^I_3\right)^{k^3_I}{\rm P}^J{\rm P}^KE^{0}_a|0\rangle\cr
&\times h^{JK}\left(\alpha_1,\{k^1_I\}\right)\bar\tau\left(\alpha_2,
\{k^2_I\}\right)\tau\left(\alpha_3,\{k^3_I\}\right)+c.c.\,\,\,.}}
\noindent An explicit derivation of the Type IIB supergravity spectrum in the pp-wave 
background was found in \refs{\mt}. In particular, the $SO(8)$ light-cone
gauge degrees of freedom of the graviton were classified according to their
$SO(4)\times SO^{\prime}(4)$ decomposition. Based on those results we can express 
various components of the graviton in terms of the mass eigenstates defined in 
\refs{\mt} as follows
\eqn\eten{\eqalign{&h_{ij}=h^{\bot}_{ij}+{1\over 8}\delta_{ij}(\rm h+\bar h)\,,\cr
&h_{i^{\prime}j^{\prime}}=h^{\bot}_{i^{\prime}j^{\prime}}-{1\over 8}
\delta_{i^{\prime}j^{\prime}}(\rm h+\bar h)\,,\cr
&h_{ij^{\prime}}=h_{j^{\prime}i}={1\over 2}(\rm h_{ij^{\prime}}+
\bar h_{ij^{\prime}})\,,}}
where $i,j=1,...,4$ and $i^{\prime},j^{\prime}=5,...,8$.
Expressed in terms of the mass eigenstates, the amplitude \ften\ becomes
\eqn\feleven{\eqalign{&A_{h\tau\bar\tau}(1,2,3)=
{1\over 3\alpha_1^2}\prod_{I=1}^8{i^{k^1_I+k^2_I+k^3_I}
\over\sqrt{k^1_I!k^2_I!k^3_I!}}\cr
\times&\biggl(\prod_{I=1}^8\langle 0|\left(a^I_1\right)^{k^1_I}
\left(a^I_2\right)^{k^2_I}\left(a^I_3\right)^{k^3_I}{\rm P}^i
{\rm P}^jE^{0}_a|0\rangle h^{\bot}_{1\,ij}\bar\tau_2\tau_3\cr
+&\prod_{I=1}^8\langle 0|\left(a^I_1\right)^{k^1_I}
\left(a^I_2\right)^{k^2_I}\left(a^I_3\right)^{k^3_I}{\rm P}^{i^{\prime}}
{\rm P}^{j^{\prime}}E^{0}_a|0\rangle h^{\bot}_{1\,i^{\prime}j^{\prime}}
\bar\tau_2\tau_3\cr
+&\prod_{I=1}^8\langle 0|\left(a^I_1\right)^{k^1_I}
\left(a^I_2\right)^{k^2_I}\left(a^I_3\right)^{k^3_I}{\rm P}^i
{\rm P}^{j^{\prime}}E^{0}_a|0\rangle {\rm h_{1\,ij^{\prime}}}
\bar\tau_2\tau_3\cr
+&\prod_{I=1}^8\langle 0|\left(a^I_1\right)^{k^1_I}
\left(a^I_2\right)^{k^2_I}\left(a^I_3\right)^{k^3_I}{\rm P}^i
{\rm P}^{j^{\prime}}E^{0}_a|0\rangle {\rm\bar h_{1\,ij^{\prime}}}
\bar\tau_2\tau_3\cr
+&{1\over 8}\prod_{I=1}^8\langle 0|\left(a^I_1\right)^{k^1_I}
\left(a^I_2\right)^{k^2_I}\left(a^I_3\right)^{k^3_I}\left({\rm P}^i{\rm P}^i-
{\rm P}^{i^{\prime}}{\rm P}^{i^{\prime}}\right)E^{0}_a|0\rangle 
{\rm h_1\bar\tau_2\tau_3}\cr
+&{1\over 8}\prod_{I=1}^8\langle 0|\left(a^I_1\right)^{k^1_I}
\left(a^I_2\right)^{k^2_I}\left(a^I_3\right)^{k^3_I}\left({\rm P}^i{\rm P}^i-
{\rm P}^{i^{\prime}}{\rm P}^{i^{\prime}}\right)
E^{0}_a|0\rangle {\rm\bar h_1\bar\tau_2\tau_3}\biggr)+c.c.\,\,\,.}}
For the last two lines in \feleven\ we can combine \fftwelve\ together with \rone\ 
and \rtwo\ to obtain
\eqn\rthree{\left({\rm P}^i{\rm P}^i-{\rm P}^{i^{\prime}}{\rm P}^{i^{\prime}}\right)
=-\alpha_1\alpha_2\alpha_3\left(\left(p^{1\,\|}_{+}+p^{2\,\|}_{+}-p^{3\,\|}_{+}\right)-
\left(p^{1\,\bot}_{+}+p^{2\,\bot}_{+}-p^{3\,\bot}_{+}\right)\right)}
and use the notation of \refs{\kklp} to define
\eqn\rfour{E^{\|}_{123}=p^{1\,\|}_{+}+p^{2\,\|}_{+}-p^{3\,\|}_{+},\,\,\,
E^{\bot}_{123}=p^{1\,\bot}_{+}+p^{2\,\bot}_{+}-p^{3\,\bot}_{+},}
where $\|$ means $i=1,..,4$ and $\bot$ means $i^{\prime}=5,..,8$.
Notice that the zero point energy contributions in \rthree\ from $\|$ 
and $\bot$ cancelled each other. In order to proceed with further 
computations we will need to evaluate expectation values of the type
\eqn\mone{\prod_{I=1}^8{i^{k^1_I+k^2_I+k^3_I}
\over\sqrt{k^1_I!k^2_I!k^3_I!}}\langle 0|\left(a^I_1\right)^{k^1_I}
\left(a^I_2\right)^{k^2_I}\left(a^I_3\right)^{k^3_I}{\rm P}^J
{\rm P}^KE^{0}_a|0\rangle\,,}
for both $J=K$ and $J\not =K$. Because of this distinction, we will split the
first and second lines of \feleven\ into the two cases and write the 
sums over $i,\,i^{\prime},\,j,\,j^{\prime}$ explicitly.
We can use \ffeleven\ and \fftwelve\ in combination with \rthree\ and \rfour\ 
and apply formulas (A.5)-(A.9) from Appendix A to obtain the final expression
for the graviton dilaton-axion off-shell amplitude
\eqn\ftwelve{\eqalign{&A_{h\tau\bar\tau\,\{{\{n^1_I\},\{n^2_I\};\{n^3_I\}\}}}
\left(\alpha_1,\alpha_2;\alpha_3\right)=\left(-\mu\alpha_1\alpha_2\alpha_3\right)\cr
&\times{1\over 3\alpha_1^2}\biggl[\sum_{i\not =j}
\left[G^i_1G^j_1+G^i_2G^j_2-G^i_3G^j_3\right]h^{\bot}_{1\,ij}\bar\tau_2\tau_3
\prod_{\scriptstyle I=1\atop\scriptstyle I\not =i,\,j}^8
K_{\{n^1_I,n^2_I;n^3_I\}}\left(\alpha_1,\,\alpha_2;\,\alpha_3\right)\cr
&+\sum_{i}\left[n^1_i+n^2_i-n^3_i\right]h^{\bot}_{1\,ii}\bar\tau_2\tau_3
\prod_{\scriptstyle I=1}^8K_{\{n^1_I,n^2_I;n^3_I\}}\left(\alpha_1,\,
\alpha_2;\,\alpha_3\right)\cr&+\sum_{i^{\prime}\not =j^{\prime}}
\left[G^{i^{\prime}}_1G^{j^{\prime}}_1+G^{i^{\prime}}_2G^{j^{\prime}}_2
-G^{i^{\prime}}_3G^{j^{\prime}}_3\right]h^{\bot}_{1\,i^{\prime}j^{\prime}}
\bar\tau_2\tau_3\prod_{\scriptstyle I=1\atop\scriptstyle I\not =
i^{\prime},\,j^{\prime}}^8K_{\{n^1_I,n^2_I;n^3_I\}}\left(\alpha_1,
\,\alpha_2;\,\alpha_3\right)\cr&+\sum_{i^{\prime}}\left[n^1_{i^{\prime}}
+n^2_{i^{\prime}}-n^3_{i^{\prime}}\right]h^{\bot}_{1\,i^{\prime}i^{\prime}}
\bar\tau_2\tau_3\prod_{\scriptstyle I=1}^8
K_{\{n^1_I,n^2_I;n^3_I\}}\left(\alpha_1,\,\alpha_2;\,\alpha_3\right)\cr
&+\sum_{i\,j^{\prime}}\left[G^i_1G^{j^{\prime}}_1+G^i_2G^{j^{\prime}}_2
-G^i_3G^{j^{\prime}}_3\right]{\rm h_{1\,ij^{\prime}}}\bar\tau_2\tau_3
\prod_{\scriptstyle I=1\atop\scriptstyle I\not =i,\,j^{\prime}}^8
K_{\{n^1_I,n^2_I;n^3_I\}}\left(\alpha_1,\,\alpha_2;\,\alpha_3\right)\cr
&+\sum_{i\,j^{\prime}}\left[G^i_1G^{j^{\prime}}_1+G^i_2G^{j^{\prime}}_2
-G^i_3G^{j^{\prime}}_3\right]{\rm\bar h_{1\,ij^{\prime}}}\bar\tau_2\tau_3
\prod_{\scriptstyle I=1\atop\scriptstyle I\not =i,\,j^{\prime}}^8
K_{\{n^1_I,n^2_I;n^3_I\}}\left(\alpha_1,\,\alpha_2;\,\alpha_3\right)\cr
&+{1\over 8\mu}\left[E^{\|}_{123}-E^{\bot}_{123}\right]
{\rm h_1\bar\tau_2\tau_3}\prod_{\scriptstyle I=1}^8
K_{\{n^1_I,n^2_I;n^3_I\}}\left(\alpha_1,\,\alpha_2;\,\alpha_3\right)\cr
&+{1\over 8\mu}\left[E^{\|}_{123}-E^{\bot}_{123}\right]
{\rm\bar h_1\bar\tau_2\tau_3}\prod_{\scriptstyle I=1}^8
K_{\{n^1_I,n^2_I;n^3_I\}}\left(\alpha_1,\,\alpha_2;\,\alpha_3\right)\biggr]+c.c.\,\,\,.}}
Here the following conditions on the occupation numbers must hold for the 
individual terms in \ftwelve\ to be non-zero
\eqn\rfive{\sum_{I=1}^8\left(n^3_I-n^1_I-n^2_I\right)\leq 2\,\,\,\, 
{\rm for\,\, terms\,\, in \,\,lines}\,\,1,\,3,\,5,\,{\rm and}\,6}
and
\eqn\rsix{\sum_{I=1}^8\left(n^3_I-n^1_I-n^2_I\right)\leq 0\,\,\,\, 
{\rm for\,\, terms\,\, in \,\,lines}\,\,2,\,4,\,7,\,{\rm and}\,8\,.}
In our next step we will apply on-shell conditions together with conservation
laws and show that in that case the amplitude \ftwelve\ will vanish.
Combining the conservation law $p^1_{+}+p^2_{+}=p^3_{+}$ and \rtwo\ 
we obtain the following condition on the occupation numbers
\eqn\ethirtytwo{\sum_{I=1}^8\left(n^3_I-n^1_I-n^2_I\right)=E^1_0+E^2_0-E^3_0\,.}
In further analysis we are going to use the results listed in TABLE I of section
3.4 of \mt\ containing the spectrum of bosonic physical degrees of freedom 
of Type IIB supergravity on the plane wave background. In particular, we will
be using the values of $E_0$ in order to analyse condition \ethirtytwo\
for various terms in \ftwelve\
\eqn\ethirtythree{\eqalign{&E_0\left({\rm h}\right)=0,\,\,\,\,\,\,\,\,
E_0\left({\rm h}_{ij^{\prime}}\right)=2,\,\,\,\,\,\,\,\,E_0\left(h^{\bot}_{ij}\right)=
E_0\left(h^{\bot}_{i^{\prime}j^{\prime}}\right)=4,\cr
&E_0\left(\tau\right)=E_0\left(\bar\tau\right)=4,\,\,\,\,\,\,\,\,
E_0\left(\bar{\rm h}_{ij^{\prime}}\right)=6,\,\,\,\,\,\,\,\,E_0
\left(\bar{\rm h}\right)=8\,.}}
For the terms in lines $1$ through $4$ of \ftwelve\ condition \ethirtytwo\ will read
\eqn\rseven{\sum_{I=1}^8\left(n^3_I-n^1_I-n^2_I\right)=4\,,}
which clearly violates the non-zero conditions \rfive\ and \rsix\ implying
that the terms in lines $1$ through $4$ vanish. After performing a similar 
check for the other terms and excluding all the terms that violate 
conditions \rfive\ and \rsix\ the amplitude reads
\eqn\fthirteen{\eqalign{&A_{h\tau\bar\tau\,\{{\{n^1_I\},\{n^2_I\};\{n^3_I\}\}}}
\left(\alpha_1,\alpha_2;\alpha_3\right)=
\left(-{\mu\alpha_1\alpha_2\alpha_3\over 3}\right)\cr
&\times\biggl({1\over \alpha_1^2}\sum_{i\,j^{\prime}}
\left[G^i_1G^{j^{\prime}}_1+G^i_2G^{j^{\prime}}_2
-G^i_3G^{j^{\prime}}_3\right]{\rm h}_{1\,ij^{\prime}}
\bar\tau_2\tau_3\prod_{\scriptstyle I=1\atop\scriptstyle I\not =
i,\,j^{\prime}}^8
K_{\{n^1_I,n^2_I;n^3_I\}}\left(\alpha_1,\,\alpha_2;\,\alpha_3\right)\cr
&+{1\over 8\mu\alpha_1^2}\left[E^{\|}_{123}-E^{\bot}_{123}\right]
{\rm h_1}\bar\tau_2\tau_3\prod_{\scriptstyle I=1}^8
K_{\{n^1_I,n^2_I;n^3_I\}}\left(\alpha_1,\,\alpha_2;\,\alpha_3\right)
+(\tau\leftrightarrow\bar\tau)\biggr)\,.}}
For the terms in \fthirteen, condition \ethirtytwo\ will read
\eqn\reight{\sum_{I=1}^8\left(n^3_I-n^1_I-n^2_I\right)=2\,\,\,\, 
{\rm for\,\, terms\,\, in \,\,line}\,\,1,}
and
\eqn\rnine{\sum_{I=1}^8\left(n^3_I-n^1_I-n^2_I\right)=0\,\,\,\, 
{\rm for\,\, terms\,\, in \,\,line}\,\,2\,.}
For the terms in the first line of \fthirteen\ we can 
arbitrarily choose two particular directions $i$ and 
$j^{\prime}$ for which $n^3_i-n^1_i-n^2_i=1$
and $n^3_{j^{\prime}}-n^1_{j^{\prime}}-n^2_{j^{\prime}}=1$ and apply
formula (A.10) while for the remaining six directions we will have
$n^3_I-n^1_I-n^2_I=0$ where $I\not =i,\,j^{\prime}$ and apply 
formula (A.4). For the terms in the second line we have the 
condition $n^3_I-n^1_I-n^2_I=0$ in all eight directions and can 
therefore apply formula (A.4). We will therefore have no summation 
over $i$ and $j^{\prime}$ for in the first line of \fthirteen. 
It will be proportional to
\eqn\ffourteen{\eqalign{
\left[{\alpha_1\over \alpha_3}+{\alpha_2\over \alpha_3}-
{|\alpha_3|\over \alpha_3}\right]
\left(n^1_i+n^2_i+1\right)^{\scriptstyle 1\over\scriptstyle 2}
\left(n^1_{j^{\prime}}+n^2_{j^{\prime}}+1\right)^{\scriptstyle 1
\over\scriptstyle 2}&\cr
\times\prod_{\scriptstyle I=1}^8
\sqrt{\left(n^1_I+n^2_I\right)!\over{n^1_I!\,n^2_I!}}
\left(-{\alpha_1\over\alpha_3}\right)^{\scriptstyle n^1_I\over\scriptstyle 2}
\left(-{\alpha_2\over\alpha_3}\right)^{\scriptstyle n^2_I\over\scriptstyle 2}&}}
and vanish due to the conservation law $\alpha_1+\alpha_2+\alpha_3=
\alpha_1+\alpha_2-|\alpha_3|=0$. The second line of \fthirteen\ 
proportional to 
\eqn\rthirty{\left(E_{123}^{\|}-E_{123}^{\bot}\right)=\mu
\left(\sum_{i=1}^4\left(n^3_i-n^1_i-n^2_i\right)-
\sum_{i^{\prime}=5}^8\left(n^3_{i^{\prime}}-n^1_{i^{\prime}}-
n^2_{i^{\prime}}\right)\right)}
will also vanish because $\,n^3_I-n^1_I-n^2_I=0\,$ for all $\,I=1,...,8$.
Therefore, the on-shell graviton dilaton-axion quantum mechanical amplitude
in the pp-wave background vanishes at tree level. As a result of this 
analysis we have to modify conditions \rfive\ and \rsix\ by replacing
the $\le$ with $<$ for the off-shell amplitude \ftwelve\ to be non-zero.
\vskip 2in
$ $
\newsec{Graviton axion-dilaton coupling in pp-wave background from 
Type IIB supergavity in light-cone gauge}
In this section we will calculate the graviton dilaton-axion cubic
interaction vertex in the pp-wave background starting from the Type IIB
supergravity action. We will see that much of the analysis of the
previous section will be carried over to this section.
The relevant piece of the Type IIB action is
\eqn\eone{S_{IIB}=-{1\over{2\kappa^{2}}}\int d^{10}x \sqrt {-g}\,{{g^{\mu \nu}
\partial_\mu \tau \partial_\nu \bar \tau}\over{2({\rm Im} \tau)^2}}\,,}
where again
\eqn\etwo{\tau(x)=\chi(x)+i e^{-\phi(x)}}
is a combination of the dilaton and axion.
Expanding the dilaton-axion field around $\phi=0,\, \chi=0$ as $\tau(x)
=i+2\kappa\tau^{\prime}(x)$
and expanding the metric around the pp-wave background as $g_{\mu \nu}(x)
=\tilde g_{\mu \nu}(x)+2 \kappa h_{\mu \nu}(x)$, we obtain from
\eone\ the following cubic vertex
\eqn\ethree{S_{h \tau \bar \tau}=2 \kappa\int d^{10}x \sqrt {-\tilde g}\,
\tilde g^{\lambda \mu} \tilde g^{\rho \nu} h_{\lambda \rho} \partial_\mu \tau 
\partial_\nu \bar \tau\,,}
where we suppressed the prime. Following Metsaev and Tseytlin \refs{\mt} we impose the 
light-cone gauge conditions\foot{This is slightly different 
from \refs{\mt} since we are using a different metric convention.}\ 
on $h_{\mu \nu}$
\eqn\efour{h_{--}=0,\,\,\,\,\,\, h_{- \mu}=0,\,\,\,\,\,\, h_{+ I}={2 \over 
\partial_{-}}\partial_{J}h_{JI},\,\,\,\,\,\, h_{++}={4 \over \partial_{-}}
\partial_{I}\partial_{J}h_{IJ},\,\,\,\,\,\, h_{II}=0\,,}
and substitute for the background metric
\eqn\efive{\tilde g^{++}={1 \over 4}\mu^2 x_I^2, \,\,\,\,\,\, \tilde g^{+-}=
\tilde g^{-+}=-{1 \over 2}, \,\,\,\,\,\, \tilde g^{IJ}=\delta^{IJ}\,,}
where $I,J=1,...,8$ and the determinant of the background metric $\tilde g=-4$. 
We then obtain from \ethree\ the following expression for the cubic 
graviton dilaton-axion interaction in the light-cone gauge
\eqn\esix{\eqalign{S_{lc}& = 4 \kappa\int dx^{+}dx^{-}d^8x \biggl[ \left({1\over(
\partial_{-})^2}\partial_I \partial_J h_{IJ}\right)\partial_{-}\bar\tau
\partial_{-}\tau
- \left({1\over \partial_{-}}\partial_J h_{JI}\right)\partial_{-}\bar\tau
\partial_{I}\tau \cr
&- \left({1\over \partial_{-}}\partial_J h_{JI}\right)\partial_{I}\bar\tau
\partial_{-}\tau+ h_{IJ}\partial_{I}\bar\tau
\partial_{J}\tau \biggr]\,.}}
Notice that the result \esix\ is $\mu$ independent and is exactly the same as 
in flat space ($\mu=0$). The same feature was found in \refs{\kklp} where 
as a functional of classical fields, the three-scalar cubic interaction 
Hamiltonian in the light-cone gauge on the pp-wave was found to be identical to 
that in flat space. However, as the authors of \refs{\kklp} have pointed out, the 
quantum mechanical amplitudes on the pp-wave will have an explicit $\mu$ dependence 
coming from the frequencies of harmonic oscillator modes. The Fock spaces for 
the flat and the pp-wave backgrounds are very different. In one case we have a 
collection of free particles, in the other case we have bound states confined by 
the gravitational potential well and described by the harmonic oscillator wave 
functions. Since the $p_{+}$ and $p_{-}$ are conserved in the pp-wave,
we will Fourier transform the fields in the light-cone directions $x^{-}$ 
and $x^{+}$ as follows
\eqn\eseven{\eqalign{h_{IJ}(x^{-},x^{+};\vec x)& = {1 \over {2 \pi}}\int dp_{+}
\int{d \alpha \over \sqrt{|\alpha|}}h_{IJ}(\alpha,p_{+};\vec x)
e^{-i(\alpha x^{-}+p_{+}x^{+})}\,,\cr
\tau(x^{-},x^{+};\vec x) &= {1 \over {2 \pi}}\int dp_{+}
\int{d \alpha \over \sqrt{|\alpha|}}\tau(\alpha,p_{+};\vec x)
e^{-i(\alpha x^{-}+p_{+}x^{+})}\,,\cr
\bar \tau(x^{-},x^{+};\vec x)& = {1 \over {2 \pi}}\int dp_{+}
\int{d \alpha \over \sqrt{|\alpha|}}\bar\tau(\alpha,p_{+};\vec x)
e^{-i(\alpha x^{-}+p_{+}x^{+})}\,,\cr}}
where $\alpha\equiv 2p_{-}$, and  obtain from \esix
\eqn\eeight{\eqalign{S_{lc}& = {2\kappa\over3}{1\over {2\pi}}
\int dp^1_{+}dp^2_{+}dp^3_{+} 
\int{d \alpha_1 \over \sqrt{|\alpha_1|}}{d \alpha_2\over \sqrt{|\alpha_2|}}
{d \alpha_3\over \sqrt{|\alpha_3|}}\int d^8x\delta(p^1_{+}+p^2_{+}+p^3_{+})\cr
&\times\delta(\alpha_1+\alpha_2+\alpha_3)\biggl[{{\alpha_2 \alpha_3}\over
{\alpha^2_1}}\left(\partial_I \partial_J h_{IJ}(\alpha_1,p^1_{+};\vec x)\right)\bar\tau
(\alpha_2,p^2_{+};\vec x)\tau(\alpha_3,p^3_{+};\vec x)\cr
&-{\alpha_2\over \alpha_1}\left(\partial_J h_{JI}(\alpha_1,p^1_{+};\vec x)\right)
\bar\tau(\alpha_2,p^2_{+};\vec x)\partial_{I}\tau(\alpha_3,p^3_{+};\vec x) \cr
&-{\alpha_3\over \alpha_1}\left(\partial_J h_{JI}(\alpha_1,p^1_{+};\vec x)\right)
\left(\partial_{I}\bar\tau(\alpha_2,p^2_{+};\vec x)\right)\tau(\alpha_3,p^3_{+};\vec x)
\cr&+ h_{IJ}(\alpha_1,p^1_{+};\vec x)\left(\partial_{I}\bar\tau(\alpha_2,
p^2_{+};\vec x)\right)\partial_{J}\tau(\alpha_3,p^3_{+};\vec x)\biggr]+c.c.\,.\cr}}
After integrating by parts and using the conservation law 
$\alpha_1+\alpha_2+\alpha_3=0$ we finally get
\eqn\enine{\eqalign{S_{lc}& = {2\kappa\over3}{1\over{2\pi}}
\int dp^1_{+}dp^2_{+}dp^3_{+} 
\int{d \alpha_1 \over \sqrt{|\alpha_1|}}{d \alpha_2\over \sqrt{|\alpha_2|}}
{d \alpha_3\over \sqrt{|\alpha_3|}}\int d^8x\delta(p^1_{+}+p^2_{+}+p^3_{+})\cr
&\times\delta(\alpha_1+\alpha_2+\alpha_3)(\alpha_1 \alpha_2 \alpha_3)\cr
&\times\biggl[{1\over
\alpha^2_1}\biggl({1\over \alpha_1}\left(\partial_I \partial_J h_{IJ}(\alpha_1,p^1_{+};
\vec x)\right)\bar\tau(\alpha_2,p^2_{+};\vec x)\tau(\alpha_3,p^3_{+};\vec x)\cr
&+{1\over \alpha_2}h_{IJ}(\alpha_1,p^1_{+};\vec x)\left(\partial_I\partial_J
\bar\tau(\alpha_2,p^2_{+};\vec x)\right)\tau(\alpha_3,p^3_{+};\vec x) \cr
&+{1\over\alpha_3}h_{IJ}(\alpha_1,p^1_{+};\vec x)\bar\tau(\alpha_2,p^2_{+};\vec x)
\partial_I\partial_J\tau(\alpha_3,p^3_{+};\vec x)\biggr)\biggr]+c.c.}}
\noindent Just as we did in the previous section, we will use the
results of \refs{\mt} to express various components of the graviton
in terms of the mass eigenstates \eten. Then the action becomes
\eqn\eeleven{\eqalign{S_{lc}& = {2\kappa\over3}{1\over {2\pi}}
\int dp^1_{+}dp^2_{+}dp^3_{+} 
\int{d \alpha_1 \over \sqrt{|\alpha_1|}}{d \alpha_2\over \sqrt{|\alpha_2|}}
{d \alpha_3\over \sqrt{|\alpha_3|}}\int d^8x\delta(p^1_{+}+p^2_{+}+p^3_{+})\cr
&\times\delta(\alpha_1+\alpha_2+\alpha_3)(\alpha_1 \alpha_2 \alpha_3)\cr
&\times{1\over \alpha^2_1}\biggl[{1\over \alpha_1}\left(\partial_i 
\partial_j h^{\bot}_{1\,ij}\right)\bar\tau_2\tau_3+{1\over \alpha_2}
h^{\bot}_{1\,ij}\left(\partial_i\partial_j\bar\tau_2\right)\tau_3
+{1\over\alpha_3}h^{\bot}_{1\,ij}\bar\tau_2\partial_i\partial_j\tau_3\cr
&+{1\over \alpha_1}\left(\partial_{i^{\prime}}\partial_{j^{\prime}} 
h^{\bot}_{1\,i^{\prime}j^{\prime}}\right)\bar\tau_2\tau_3+{1\over\alpha_2}
h^{\bot}_{1\,i^{\prime}j^{\prime}}\left(\partial_{i^{\prime}}\partial_{j^{\prime}}
\bar\tau_2\right)\tau_3+{1\over\alpha_3}h^{\bot}_{1\,i^{\prime}j^{\prime}}\bar\tau_2
\partial_{i^{\prime}}\partial_{j^{\prime}}\tau_3\cr
&+{1\over \alpha_1}\left(\partial_i\partial_{j^{\prime}} 
{\rm h}_{1\,ij^{\prime}}\right)\bar\tau_2\tau_3+{1\over\alpha_2}
{\rm h}_{1\,ij^{\prime}}\left(\partial_i\partial_{j^{\prime}}
\bar\tau_2\right)\tau_3+{1\over\alpha_3}{\rm h}_{1\,ij^{\prime}}\bar\tau_2
\partial_i\partial_{j^{\prime}}\tau_3\cr
&+{1\over \alpha_1}\left(\partial_i\partial_{j^{\prime}} 
\bar{\rm h}_{1\,ij^{\prime}}\right)\bar\tau_2\tau_3+{1\over\alpha_2}
\bar{\rm h}_{1\,ij^{\prime}}\left(\partial_i\partial_{j^{\prime}}
\bar\tau_2\right)\tau_3+{1\over\alpha_3}\bar{\rm h}_{1\,ij^{\prime}}\bar\tau_2
\partial_i\partial_{j^{\prime}}\tau_3\cr
&+{1\over {8\alpha_1}}\left((\partial_i^2-\partial_{i^{\prime}}^2)
{\rm h_1}\right)\bar\tau_2\tau_3+{1\over{8\alpha_2}}
{\rm h_1}\left((\partial_i^2-\partial_{i^{\prime}}^2)
\bar\tau_2\right)\tau_3+{1\over{8\alpha_3}}{\rm h}_1\bar\tau_2
(\partial_i^2-\partial_{i^{\prime}}^2)\tau_3\cr
&+{1\over {8\alpha_1}}\left((\partial_i^2-\partial_{i^{\prime}}^2)
\bar{\rm h}_1\right)\bar\tau_2\tau_3+{1\over{8\alpha_2}}
\bar{\rm h}_1\left((\partial_i^2-\partial_{i^{\prime}}^2)
\bar\tau_2\right)\tau_3+{1\over{8\alpha_3}}\bar{\rm h}_1\bar\tau_2
(\partial_i^2-\partial_{i^{\prime}}^2)\tau_3\biggr]\cr
&+c.c.\,.}}
In further calculations we assume that state 3 is incoming and states 1 and 2 are 
outgoung. This is easily achieved by relabelling $\alpha_3 \rightarrow -\alpha_3$
and $p^3_{+} \rightarrow -p^3_{+}$.
The conservation laws will then read $\alpha_1+\alpha_2-\alpha_3=0$ and 
$p^1_{+}+p^2_{+}-p^3_{+}=0$ where all $\alpha$'s and $p_{+}$'s are now positive definite.
The dynamics of the fields in the transverse directions is governed by the light-cone 
Hamiltonian  \refs{\svone} given by
\eqn\etwelve{{\rm P}_{+}=-{1\over{\alpha}}\partial_I^2+{\mu^2 {\alpha}
\over 4}x_I^2+(E_0 -4)\mu\,.}
This can be split into two contributions from the two $SO(4)$ directions as
\eqn\ethirteen{{\rm P}_{+}={\rm P}_{+}^{\|}+{\rm P}_{+}^{\bot},}
where we have defined
\eqn\ethirteen{\eqalign{&{\rm P}_{+}^{\|}=-{1\over\alpha}\partial_i^2
+{\mu^2 \alpha\over 4}x_i^2+\left({E_0\over2} -2\right)\mu,\cr
&{\rm P}_{+}^{\bot}=-{1\over\alpha}\partial^2_{i^{\prime}}+{\mu^2 \alpha\over 4}
{x_{i^{\prime}}^2+\left({E_0\over 2} -2\right)\mu.}}}
Following the line of argument presented in \refs{\kklp} we notice that if we
use the conservation law $\alpha_1+\alpha_2-\alpha_3=0$, we can insert 
into the action terms proportional to
\eqn\efourteen{{\mu^2\over 4}(\alpha_1+\alpha_2-\alpha_3)(x^2_i-x^2_{i^{\prime}})}
without changing it. This allows us to combine ${\rm h}\bar\tau\tau$ terms as
\eqn\efifteen{\eqalign{
&-{1\over {8\alpha_1}}\left((\partial_i^2-\partial_{i^{\prime}}^2)
{\rm h_1}\right)\bar\tau_2\tau_3-{1\over{8\alpha_2}}
{\rm h_1}\left((\partial_i^2-\partial_{i^{\prime}}^2)
\bar\tau_2\right)\tau_3\cr
&+{1\over{8\alpha_3}}{\rm h}_1\bar\tau_2
(\partial_i^2-\partial_{i^{\prime}}^2)\tau_3=
{1\over 8}\left(\left({\rm P}_{+}^{\|}-{\rm P}_{+}^{\bot}\right)
{\rm h_1}\right)\bar\tau_2\tau_3\cr
&+{1\over 8}{\rm h_1}\left(\left({\rm P}_{+}^{\|}-{\rm P}_{+}^{\bot}\right)
\bar\tau_2\right)\tau_3-{1\over 8}{\rm h}_1\bar\tau_2
\left({\rm P}_{+}^{\|}-{\rm P}_{+}^{\bot}\right)\tau_3\cr
&={1\over 8}\left(\left(p^{1\,\|}_{+}+p^{2\,\|}_{+}-p^{3\,\|}_{+}\right)
-\left(p^{1\,\bot}_{+}+p^{2\,\bot}_{+}-p^{3\,\bot}_{+}\right)\right)
{\rm h_1}\bar\tau_2\tau_3\,,}}
where we also used the fact that the zero point energy contributions from the two
$SO(4)$ directions cancel. Similarly, we can combine $\bar {\rm h}
\bar\tau\tau$ terms to obtain the following expression for the action
\eqn\esixteen{\eqalign{S_{lc}& = {2\kappa\over3}{1\over {2\pi}}
\int dp^1_{+}dp^2_{+}dp^3_{+} 
\int{{d\alpha_1 d\alpha_2 d\alpha_3}\over\sqrt{\alpha_1\alpha_2\alpha_3}}
\int d^8x\delta(p^1_{+}+p^2_{+}-p^3_{+})\cr
&\times\delta(\alpha_1+\alpha_2-\alpha_3)(\alpha_1 \alpha_2 \alpha_3)\cr
&\times{1\over \alpha^2_1}\biggl[-{1\over \alpha_1}\left(\partial_i 
\partial_j h^{\bot}_{1\,ij}\right)\bar\tau_2\tau_3-{1\over \alpha_2}
h^{\bot}_{1\,ij}\left(\partial_i\partial_j\bar\tau_2\right)\tau_3
+{1\over\alpha_3}h^{\bot}_{1\,ij}\bar\tau_2\partial_i\partial_j\tau_3\cr
&-{1\over \alpha_1}\left(\partial_{i^{\prime}}\partial_{j^{\prime}} 
h^{\bot}_{1\,i^{\prime}j^{\prime}}\right)\bar\tau_2\tau_3-{1\over\alpha_2}
h^{\bot}_{1\,i^{\prime}j^{\prime}}\left(\partial_{i^{\prime}}\partial_{j^{\prime}}
\bar\tau_2\right)\tau_3+{1\over\alpha_3}h^{\bot}_{1\,i^{\prime}j^{\prime}}\bar\tau_2
\partial_{i^{\prime}}\partial_{j^{\prime}}\tau_3\cr
&-{1\over \alpha_1}\left(\partial_i\partial_{j^{\prime}} 
{\rm h}_{1\,ij^{\prime}}\right)\bar\tau_2\tau_3-{1\over\alpha_2}
{\rm h}_{1\,ij^{\prime}}\left(\partial_i\partial_{j^{\prime}}
\bar\tau_2\right)\tau_3+{1\over\alpha_3}{\rm h}_{1\,ij^{\prime}}\bar\tau_2
\partial_i\partial_{j^{\prime}}\tau_3\cr
&-{1\over \alpha_1}\left(\partial_i\partial_{j^{\prime}} 
\bar{\rm h}_{1\,ij^{\prime}}\right)\bar\tau_2\tau_3-{1\over\alpha_2}
\bar{\rm h}_{1\,ij^{\prime}}\left(\partial_i\partial_{j^{\prime}}
\bar\tau_2\right)\tau_3+{1\over\alpha_3}\bar{\rm h}_{1\,ij^{\prime}}\bar\tau_2
\partial_i\partial_{j^{\prime}}\tau_3\cr
&+{1\over 8}\left(\left(p^{1\,\|}_{+}+p^{2\,\|}_{+}-p^{3\,\|}_{+}\right)
-\left(p^{1\,\bot}_{+}+p^{2\,\bot}_{+}-p^{3\,\bot}_{+}\right)\right)
{\rm h_1}\bar\tau_2\tau_3\cr
&+{1\over 8}\left(\left({p^1}_{+}^{\|}+{p^2}_{+}^{\|}-{p^3}_{+}^{\|}\right)
-\left(p^{1\,\bot}_{+}+p^{2\,\bot}_{+}-p^{3\,\bot}_{+}\right)\right)
{\bar{\rm h}_1}\bar\tau_2\tau_3\biggr]+c.c.\,.}}
Corresponding to the Hamiltonian \etwelve\ are eight-dimensional 
harmonic oscillator wave functions $\psi_{\vec k}\left(\sqrt{{\mu\alpha}
\over 2}\vec x\right)$ written explicitly in Appendix B. They form
a complete basis. It is natural to expand our interacting fields
in such a basis
\eqn\eseventeen{\eqalign{&h^{\bot}_{ij}\left(\alpha,p_{+};\vec x\right)=\sum_{\vec k}
h^{\bot}_{ij}\left(\alpha,p_{+};\vec k\right)\psi_{\vec k}\left(\sqrt{{\mu\alpha}
\over 2}\vec x\right)\,,\cr
&h^{\bot}_{i^{\prime}j^{\prime}}\left(\alpha,p_{+};\vec x\right)=\sum_{\vec k}
h^{\bot}_{i^{\prime}j^{\prime}}\left(\alpha,p_{+};\vec k\right)\psi_{\vec k}
\left(\sqrt{{\mu\alpha}\over 2}\vec x\right)\,,\cr
&{\rm h}_{ij^{\prime}}\left(\alpha,p_{+};\vec x\right)=\sum_{\vec k}
{\rm h}_{ij^{\prime}}\left(\alpha,p_{+};\vec k\right)\psi_{\vec k}
\left(\sqrt{{\mu\alpha}\over 2}\vec x\right)\,,\cr
&{\rm h}\left(\alpha,p_{+};\vec x\right)=\sum_{\vec k}
{\rm h}\left(\alpha,p_{+};\vec k\right)\psi_{\vec k}
\left(\sqrt{{\mu\alpha}\over 2}\vec x\right)\,,\cr
&{\tau}\left(\alpha,p_{+};\vec x\right)=\sum_{\vec k}
{\tau}\left(\alpha,p_{+};\vec k\right)\psi_{\vec k}
\left(\sqrt{{\mu\alpha}\over 2}\vec x\right)\,,}}
and similarly for ${\bar{\rm h}}_{ij^{\prime}}$, $\bar{\rm h}$ and $\bar\tau$.
The following analysis will be analogous to that of the previous section but 
now we will use the properties of the eight-dimensional harmonic oscillator
wave functions given in Appendix B in place of the Fock space amplitudes of
Appendix A. If we were now to calculate an off-shell interaction vertex
from \esixteen\ using expansions \eseventeen\ we would have to use 
the most general case expression for (B.5)  given in \refs{\dbsj} 
and formulas (B.9)-(B.11). We state in Appendix A, that the general case formula 
for the string calculation (A.2) will reduce to (B.5) up to a normalization 
factor if we use the Neumann coefficients for the supergravity vertex given 
in \refs{\svone}. Likewise, certain products of the string formulas (A.6)-(A.8) 
will reduce to the general case for (B.9)-(B.11). We can therefore conclude that 
up to a normalization factor, the off-shell interaction amplitude \ftwelve\ 
containing the $\alpha^{\prime}$ corrections will simply reduce for 
$\alpha^{\prime}\rightarrow 0$ to an expression that we could also have obtained 
from \esixteen\ which would have no string corrections. Although further 
analysis is almost identical to the one we carried out at the end the previous 
section, we will nevertheless include it for the purpose of completeness.
Once we insert the expansions \eseventeen\ into the action \esixteen, we will
apply conditions (B.6) and (B.12) in combination with \ethirtytwo\ and \ethirtythree\
to see which terms survive. For instance, let us consider the
first nine terms in \esixteen\ of the form $h^{\bot}_{ij}\bar\tau\tau$ and 
$h^{\bot}_{i^{\prime}j^{\prime}}\bar\tau\tau$ with various second derivatives. 
For all those terms \ethirtytwo\ reads
\eqn\ethirtyfour{\sum_{I=1}^8\left(k^3_I-k^1_I-k^2_I\right)=4.}
After we substitute \eseventeen\ into \esixteen\ all those terms will result 
in integrals of type (B.9)-(B.11) and will automatically vanish because 
\ethirtyfour\ implies (B.12).
Performing a similar analysis for the other terms and dropping all those that vanish 
the action now reads
\eqn\esixteena{\eqalign{S_{lc}& = {2\kappa\over3}{1\over {2\pi}}
\int dp^1_{+}dp^2_{+}dp^3_{+} 
\int{{d\alpha_1 d\alpha_2 d\alpha_3}\over\sqrt{\alpha_1\alpha_2\alpha_3}}
\int d^8x\delta(p^1_{+}+p^2_{+}-p^3_{+})\cr
&\times\delta(\alpha_1+\alpha_2-\alpha_3)(\alpha_1 \alpha_2 \alpha_3)\cr
&\times\biggl[{1\over \alpha^2_1}
\left(-{1\over \alpha_1}\left(\partial_i\partial_{j^{\prime}} 
{\rm h}_{1\,ij^{\prime}}\right)\bar\tau_2\tau_3-{1\over\alpha_2}
{\rm h}_{1\,ij^{\prime}}\left(\partial_i\partial_{j^{\prime}}
\bar\tau_2\right)\tau_3+{1\over\alpha_3}{\rm h}_{1\,ij^{\prime}}\bar\tau_2
\partial_i\partial_{j^{\prime}}\tau_3\right)\cr
&+{1\over {8 \alpha^2_1}}\left(E_{123}^{\|}-E_{123}^{\bot}\right)
{\rm h_1}\bar\tau_2\tau_3\biggr]+\left(\tau\leftrightarrow\bar\tau\right)\,.}}
Here we again used the notation of \refs{\kklp} to define
\eqn\energ{E^{\|}_{123}=p^{1\,\|}_{+}+p^{2\,\|}_{+}-p^{3\,\|}_{+},\,\,\,
E^{\bot}_{123}=p^{1\,\bot}_{+}+p^{2\,\bot}_{+}-p^{3\,\bot}_{+}.}
We see that all the remaining terms constitute special cases
described in Appendix B. Namely, for the three terms in \esixteena\ 
containing second derivatives, condition \ethirtytwo\ reads
\eqn\ethirtytwob{\sum_{I=1}^8\left(k^3_I-k^1_I-k^2_I\right)=2,}
while for the last terms with no derivatives it reads
\eqn\ethirtytwoc{\sum_{I=1}^8\left(k^3_I-k^1_I-k^2_I\right)=0.}
We notice immediately that \ethirtytwob\ implies (B.13) so we can use
(B.14)-(B.16) to evaluate the integrals (B.9)-(B.11) appearing
in the terms of \esixteena\ containing second derivatives. Since 
\ethirtytwoc\ implies (B.7) we can use (B.8) to evaluate the last terms.
The sum of the second derivative terms in \esixteena\ will then be 
proportional to
\eqn\eform{\eqalign{&\left(-{1\over\alpha_1}I_1-{1\over\alpha_2}I_2
+{1\over\alpha_3}I_3\right)\cr
&={\mu\over {4\alpha_3}}\left(-\alpha_1-\alpha_2+\alpha_3\right)
\left(k^1_i+k^2_i+1\right)^{\scriptstyle 1\over\scriptstyle 2}
\left(k^1_{j^{\prime}}+k^2_{j^{\prime}}+1\right)^
{\scriptstyle 1\over\scriptstyle 2}\cr
&\times F_{\{\vec k_1,\vec k_2;\vec k_1+\vec k_2\}}(\alpha_1,\,\alpha_2;\,
\alpha_3)\,,}}
and will vanish due the conservation law $\alpha_1+\alpha_2-\alpha_3=0$.
The last term in \esixteena\ is proportional to
\eqn\ena{\left(E_{123}^{\|}-E_{123}^{\bot}\right)=\mu
\left(\sum_{i=1}^4\left(k^3_i-k^1_i-k^2_i\right)-
\sum_{i^{\prime}=5}^8\left(k^3_{i^{\prime}}-k^1_{i^{\prime}}-
k^2_{i^{\prime}}\right)\right)\,,}
where the zero point energy contributions cancelled each other
and since $k^1_I+k^2_I=k^3_I$ for $I=1,...,8\,$, the term
proportional to $\left(E_{123}^{\|}-E_{123}^{\bot}\right)$
vanishes because individual terms inside the sums in \ena\ are zero.
\vskip20pt
\newsec{Conclusion}
In this paper we explicitly calculated the graviton dilaton-axion three-point
vertex in the light-cone gauge in the pp-wave background. In section 2 we employed 
the light-cone string field theory formalism to obtain the off-shell vertex 
containing the stringy $\alpha^{\prime}$ corrections. Through a careful
analysis we showed that the vertex vanishes when all the particles are on-shell 
and the conservation laws are imposed. In section 3 we approached the same
problem from the low energy limit and expanded a particular sector of
the Type IIB supergravity action around the pp-wave background in the 
light-cone gauge. We then analysed the supergravity graviton dilaton-axion vertex 
using the properties of the eight-dimensional harmonic oscillator wave 
functions and showed that the interaction vertex vanishes
when the conservation laws are combined with the on-shell conditions.

The authors of \refs{\dbsj} have investigated a possibility of
the S-matrix formulation for scalar field theories as well as a 
gauge theory coupled to scalars on the plane-wave background.
In section (3.2.1) of  \refs{\dbsj} it was shown that for gauge theory 
coupled to scalars, the corresponding interaction vertex for the on-shell 
``photon'' exchange vanishes after some non-trivial cancellations. 
This result was used to argue that the propagator for the
exchanged particle in a four-point amplitude will never blow-up since
the exchanged particle will always be off-shell. This, together with the condition 
of convergence for the four-point amplitudes allowed them to conclude that 
the S-matrix interpretation for field theories on the plane wave background 
exists, at least at the tree level.

Our result is in the spirit of \refs{\dbsj} and we can similarly 
conclude that in four-point amplitudes involving scalars coupled to gravity on 
the pp-wave background, the propagator of the exchanged particle will
never blow up and the potentially dangerous ``graviton'' will always be off-shell.
Moreover, our result is true not only in the case of a field theory such as
Type IIB supergravity (section 3), but also in the case of the 
full Type IIB string theory (section 2).
\vskip20pt
\appendix{A}{Three-Particle Bosonic Amplitudes}
\noindent For our  computations we need to evaluate the expectation value given by
\eqn\kone{K_{\{n^1_I,n^2_I;n^3_I\}}\left(\alpha_1,\,\alpha_2;\,\alpha_3\right)=
{i^{n^1_I+n^2_I+n^3_I}\over\sqrt{n^1_I!n^2_I!n^3_I!}}
\langle 0|\left(a^I_1\right)^{n^1_I}\left(a^I_2\right)^{n^2_I}
\left(a^I_3\right)^{n^3_I}E^{0}_a|0\rangle\,,}
where $E^{0}_a$ is given in \feseven.
Taking into account the fact that $\overline N_{33}=0$, after a careful analysis 
we obtain the following formula for the general case 
\eqn\ktwo{\eqalign{K_{\{n^1_I,n^2_I;n^3_I\}}
\left(\alpha_1,\,\alpha_2;\,\alpha_3\right)=i^{n^1_I+n^2_I+n^3_I}
\sqrt{n^1_I!\,n^2_I!\,n^3_I!\over{2^{n^1_I+n^2_I+n^3_I}}}\sum_{l_1=0}^{n^1_I}
\sum_{l_2=0}^{n^2_I}
{\left(\overline N_{11}\right)^{\scriptstyle n^1_I-l_1\over\scriptstyle 2}
\over{\left(n^1_I-l_1\over 2\right)!}}&\cr
\times{\left(\overline N_{22}\right)^{\scriptstyle n^2_I-l_2\over\scriptstyle 2}
\left(2\overline N_{13}\right)^{\scriptstyle n^3_I+l_1-l_2\over\scriptstyle 2}
\left(2\overline N_{23}\right)^{\scriptstyle n^3_I+l_2-l_1\over\scriptstyle 2}
\left(2\overline N_{12}\right)^{\scriptstyle l_1+l_2-n^3_I\over\scriptstyle 2}
\over\left(n^2_I-l_2\over 2\right)!\left(n^3_I+l_1-l_2\over 2\right)!
\left(n^3_I+l_2-l_1\over 2\right)!\left(l_1+l_2-n^3_I\over 2\right)!}&
\,,}}
where ${n^3_I+l_1+l_2\over 2}$ is an integer and ${n^3_I+l_1+l_2\over 2}\ge{\rm max}
\{n^3_I,l_1,l_2\}$ and both $n^1_I-l_1$ and $n^2_I-l_2$ must be even. 
Therefore $K_{\{n^1_I,n^2_I;n^3_I\}}\left(\alpha_1,\,\alpha_2;\,\alpha_3\right)$ is 
non-zero only for $n^3_I-n^1_I-n^2_I\leq 0$.
If we use the Neumann coefficients for the supergravity vertex
given in \refs{\svone}, expression \ktwo\ will reduce to formula (A.5) 
given in Appendix A of \refs{\dbsj} up to a factor of
$\sqrt\pi\,\sqrt{2^{\scriptstyle n^1_I+n^2_I+n^3_I}\left(n^1_I!\,n^2_I!\,n^3_I!\right)}$.
Formula (B.5) from the next section will represent precisely such case.
For a special case when $n^3_I-n^1_I-n^2_I=0$ formula \ktwo\ will reduce to
\eqn\kthree{K_{\{n^1_I,n^2_I;n^1_I+n^2_I\}}\left(\alpha_1,\,
\alpha_2;\,\alpha_3\right)=\left(-1\right)^{n^1_I+n^2_I}
\sqrt{\left(n^1_I+n^2_I\right)!\over{n^1_I!\,n^2_I!}}
\,\overline N_{13}^{n^1_I}\,\overline N_{23}^{n^2_I}\,,}
which was derived in \refs{\lmpone}. Notice, that the Neumann coefficients
for the zero modes of the string vertex in \fenine\ coincide with those of the 
supergravity vertex $M_{13}$ and $M_{23}$ given in \refs{\svone}. Substituting
explicitly for $\overline N_{13}$ and $\overline N_{23}$ we obtain from \kthree
\eqn\kfour{K_{\{n^1_I,n^2_I;n^1_I+n^2_I\}}\left(\alpha_1,\,
\alpha_2;\,\alpha_3\right)=\sqrt{\left(n^1_I+n^2_I\right)!\over{n^1_I!\,n^2_I!}}
\left(-{\alpha_1\over\alpha_3}\right)^{\scriptstyle n^1_I\over\scriptstyle 2}
\left(-{\alpha_2\over\alpha_3}\right)^{\scriptstyle n^2_I\over\scriptstyle 2}\,,}
which coincides with formula (A.6) given in Appendix A of \refs{\dbsj} up to a 
factor of $\sqrt\pi\,2^{\scriptstyle n^1_I+n^2_I}
\sqrt{\left(n^1_I+n^2_I\right)!\,n^1_I!\,n^2_I!}$.
Another expectation value that we would like to evaluate is
\eqn\kfive{G^{\,I}_{r\,{\{n^1_I,n^2_I;n^3_I\}}}\left(\alpha_1,\,\alpha_2;\,
\alpha_3\right)={i^{n^1_I+n^2_I+n^3_I}\over\sqrt{n^1_I!n^2_I!n^3_I!}}
\langle 0|\left(a^I_1\right)^{n^1_I}\left(a^I_2\right)^{n^2_I}
\left(a^I_3\right)^{n^3_I}\left(a_r^I+a_r^{\dagger I}\right)E^{0}_a|0\rangle\,.}
Using the commutation relation in \feight\ in combination with the 
definition \kone, we obtain for 
$n^3_I-n^1_I-n^2_I\leq 1$
\eqn\ksixone{\eqalign{G^{\,I}_{1\,{\{n^1_I,n^2_I;n^3_I\}}}
\left(\alpha_1,\,\alpha_2;\,\alpha_3\right)&=
-i\sqrt{n^1_I+1}K_{\{n^1_I+1,n^2_I;n^3_I\}}
\left(\alpha_1,\,\alpha_2;\,\alpha_3\right)\cr
&+i\sqrt{n^1_I}\,K_{\{n^1_I-1,n^2_I;n^3_I\}}
\left(\alpha_1,\,\alpha_2;\,\alpha_3\right)\,,}}
\eqn\ksixtwo{\eqalign{G^{\,I}_{2\,{\{n^1_I,n^2_I;n^3_I\}}}
\left(\alpha_1,\,\alpha_2;\,\alpha_3\right)&=
-i\sqrt{n^2_I+1}K_{\{n^1_I,n^2_I+1;n^3_I\}}
\left(\alpha_1,\,\alpha_2;\,\alpha_3\right)\cr
&+i\sqrt{n^2_I}\,K_{\{n^1_I,n^2_I-1;n^3_I\}}
\left(\alpha_1,\,\alpha_2;\,\alpha_3\right)\,,}}
\eqn\ksixthree{\eqalign{G^{\,I}_{3\,{\{n^1_I,n^2_I;n^3_I\}}}
\left(\alpha_1,\,\alpha_2;\,\alpha_3\right)&=
-i\sqrt{n^3_I+1}K_{\{n^1_I,n^2_I;n^3_I+1\}}
\left(\alpha_1,\,\alpha_2;\,\alpha_3\right)\cr
&+i\sqrt{n^3_I}\,K_{\{n^1_I,n^2_I;n^3_I-1\}}
\left(\alpha_1,\,\alpha_2;\,\alpha_3\right)\,,}}
and
\eqn\ksixone{G^{\,I}_{r\,{\{n^1_I,n^2_I;n^3_I\}}}
\left(\alpha_1,\,\alpha_2;\,\alpha_3\right)=0,\,\,
{\rm if}\,\,\,\,n_I^3-n_I^1-n_I^2>1\,.}
For a special case when $n^3_I-n^1_I-n^2_I=1$, \kfive\ becomes
\eqn\kseven{\eqalign{G^{\,I}_{r\,{\{n^1_I,n^2_I;n^1_I+n^2_I+1\}}}
\left(\alpha_1,\,\alpha_2;\,\alpha_3\right)&=-i\,sign(\alpha_r)(n^1_I+n^2_I+1)^
{\scriptstyle 1\over\scriptstyle 2}
\left(-{|\alpha_r|\over\alpha_3}\right)^{\scriptstyle 1\over\scriptstyle 2}\cr
&\times K_{\{n^1_I,n^2_I;n^1_I+n^2_I\}}\left(\alpha_1,\,\alpha_2;\,\alpha_3\right)\,.}}
\vskip20pt
\appendix{B} {Integrals Involving Harmonic Oscillator Wave Functions}
Here we will list a few useful formulas and identities involving the 
eight-dimensional harmonic oscillator wave functions. This section will contain
expressions that can be easily derived based on the formulas given
in the Appendix A of \refs{\dbsj} as well as in \refs{\gr}.
\noindent
The light-cone Hamiltonian for a physical field is
\eqn\eeighteen{{\rm P}_{+}=-{1\over{\alpha}}\partial_I^2+{\mu^2 {\alpha}
\over 4}x_I^2+(E_0 -4)\mu\,,}
where $(E_0 -4)\mu$ is a contribution to the zero point energy coming from
the fermionic zero modes.
The corresponding wave function is 
\eqn\enineteen{\psi_{\vec k}\left(\sqrt{{\mu\alpha}\over 2}\vec x\right)=
\prod_{I=1}^{8}\psi_{k_I}\left(\sqrt{{\mu\alpha}\over 2}x_I\right)\,,}
where
\eqn\etwenty{\psi_{k_I}\left(\sqrt{\mu\alpha\over 2}x_I\right)=\left({\alpha\mu}
\over{2\pi}\right)^{\scriptstyle 1\over\scriptstyle 4}{1\over\sqrt{2^{k_I}k_I!}}
e^{-\mu\alpha x_{I}^2/4}
H_{k_I}\left(\sqrt{\mu\alpha\over 2}x_I\right)\,,}
with the energy
\eqn\etwentyone{p_{+}=\mu\left(\sum_{I=1}^8k_{I}+E_0\right)\,.}
Following the notation of \refs{\dbsj} we define
\eqn\etwentytwo{\eqalign{&F_{\{\vec k_1,\vec k_2;\vec k_3\}}(\alpha_1,\,
\alpha_2;\,\alpha_3)=\prod_{I=1}^8F_{\{k^1_I,k^2_I;k^3_I\}}(\alpha_1,\,
\alpha_2;\,\alpha_3)\,,\cr
&F_{\{k^1_I,k^2_I;k^3_I\}}(\alpha_1,\,\alpha_2;\,\alpha_3)\cr
&=\int\psi_{k^1_I}\left(\sqrt{\mu\alpha_1\over 2}
x_I\right)\psi_{k^2_I}\left(\sqrt{\mu\alpha_2\over 2}x_I\right)\psi_{k^3_I}
\left(\sqrt{\mu\alpha_3\over 2}x_I\right)dx_I\,,}}
where $\alpha_3=\alpha_1+\alpha_2$. The general expression for
 $F_{\{k^1_I,k^2_I;k^3_I\}}(\alpha_1,\,\alpha_2;\,\alpha_3)$ when
$k^3_I-k^1_I-k^2_I\leq 0$ was given in \refs{\dbsj} 
and it was found that it vanishes if $k^3_I-k^1_I-k^2_I>0$. 
Therefore we have the following condition
\eqn\etwentynine{F_{\{\vec k^1,\vec k^2;\vec k^3\}}(\alpha_1,\,\alpha_2;\,\alpha_3)
=0{,}
\,\,\,\,{\rm if}\,\,\,\,\,\sum_{I=1}^8\left(k_I^3-k_I^1-k_I^2\right)>0\,.}
Of particular interest will be the case 
when $k^3_I=k^1_I+k^2_I$. In this special case also given in \refs{\dbsj},
\eqn\etwentyfour{\sum_{I=1}^8\left(k_I^1+k_I^2-k_I^3\right)=0\,,}
\eqn\etwentyfoura{F_{\{\vec k_1,\vec k_2;\vec k_1+\vec k_2\}}(\alpha_1,\,
\alpha_2;\,\alpha_3)
=\prod_{I=1}^8\left(\mu\,\alpha_1\alpha_2\over 2\pi\alpha_3\right)^
{\scriptstyle 1\over\scriptstyle 4}
\sqrt{(k^1_I+k^2_I)!\over{k^1_I!k^2_I!}}\left(\alpha_1\over\alpha_3
\right)^{\scriptstyle k^1_I\over\scriptstyle 2}\left(\alpha_2\over\alpha_3\right)^
{\scriptstyle k^2_I\over\scriptstyle 2}\,.}
Other cases of interest will involve integrals with second derivatives of the
wave functions. The cases relevant to our calculations will involve
\eqn\etwentysix{
I_1=\prod_{I=1}^8\int\left(\partial_J\partial_K\psi_{k^1_I}\left(\sqrt{\mu\alpha_1
\over 2}
x_I\right)\right)\psi_{k^2_I}\left(\sqrt{\mu\alpha_2\over 2}x_I\right)\psi_{k^3_I}
\left(\sqrt{\mu\alpha_3\over 2}x_I\right)dx_I\,,}
\eqn\etwentyseven{
I_2=\prod_{I=1}^8\int\psi_{k^1_I}\left(\sqrt{\mu\alpha_1\over 2}
x_I\right)\left(\partial_J\partial_K\psi_{k^2_I}\left(\sqrt{\mu\alpha_2\over 2}x_I\right)
\right)\psi_{k^3_I}\left(\sqrt{\mu\alpha_3\over 2}x_I\right)dx_I\,,}
\eqn\etwentyfive{
I_3=\prod_{I=1}^8\int\psi_{k^1_I}\left(\sqrt{\mu\alpha_1\over 2}
x_I\right)\psi_{k^2_I}\left(\sqrt{\mu\alpha_2\over 2}x_I\right)\left(\partial_J\partial_K
\psi_{k^3_I}\left(\sqrt{\mu\alpha_3\over 2}x_I\right)\right)dx_I\,.}
Using formulas (A.5)-(A.7) of \refs{\dbsj} it is straightforward to show that
\eqn\ethirtyone{I_1=I_2=I_3=0,\,\,{\rm if}\,\,
\sum_{I=1}^8\left(k_I^3-k_I^1-k_I^2\right)>2\,.}
For purposes of the calculation it is important to note that \ethirtyone\ 
is true for both $J=K$ and $J\not =K$. For a special case 
\eqn\eas{\sum_{I=1}^8\left(k_I^3-k_I^1-k_I^2\right)=2\,,}
we have
\eqn\etwentysixa{
I_1={\mu\alpha_1^2\over 4\alpha_3}\left(k^1_J+k^2_J+1\right)^
{\scriptstyle 1\over\scriptstyle 2}\left(k^1_K+k^2_K+1
\right)^{\scriptstyle 1\over\scriptstyle 2}
F_{\{\vec k_1,\vec k_2;\vec k_1+\vec k_2\}}(\alpha_1,\,\alpha_2;\,\alpha_3)\,,}
\eqn\etwentysevena{
I_2={\mu\alpha_2^2\over 4\alpha_3}\left(k^1_J+k^2_J+1\right)^
{\scriptstyle 1\over\scriptstyle 2}\left(k^1_K+k^2_K+1
\right)^{\scriptstyle 1\over\scriptstyle 2}
F_{\{\vec k_1,\vec k_2;\vec k_1+\vec k_2\}}(\alpha_1,\,\alpha_2;\,\alpha_3)\,,}
\eqn\etwentyfivea{
I_3={\mu\alpha_3\over 4}\left(k^1_J+k^2_J+1\right)^{\scriptstyle 1\over\scriptstyle 2}
\left(k^1_K+k^2_K+1\right)^{\scriptstyle 1\over\scriptstyle 2}
F_{\{\vec k_1,\vec k_2;\vec k_1+\vec k_2\}}(\alpha_1,\,\alpha_2;\,\alpha_3)\,,}
where $J\not= K$ and the occupation numbers must satisfy the condition
\eqn\etwentyeight{
k^3_I=\cases{ 
k^1_I+k^2_I+1 & when $I=J$ or $I=K$ \cr
k^1_I+k^2_I & otherwise \cr}\,.}
\vskip20pt
\appendix{C}{Some $\gamma$ matrix identities}
This section contains some $SO(8)$ $\gamma$ matrix identities that we derived using
MathTensor and FeynCalc packages for Mathematica.
The gamma matrices satisfy 
\eqn\gamone{\gamma^I_{a\,\dot c}\gamma^J_{\dot c\,b}
+\gamma^J_{a\,\dot c}\gamma^I_{\dot c\,b}=2\delta^{IJ}\delta_{ab}\,,}
and
\eqn\gamtwo{\gamma^{IJ}_{a\,b}={1\over 2}\left(\gamma^I_{a\,\dot c}\gamma^J_{\dot c\,b}
-\gamma^J_{a\,\dot c}\gamma^I_{\dot c\,b}\right)\,.}
The self dual tensor $t^{IJ}_{abcd}$ is defined as follows
\eqn\gamthree{t^{IJ}_{abcd}=\gamma^{IK}_{[ab}\gamma^{JK}_{cd]}\,.}
Based on the definitions \gamone-\gamthree, we have derived the following two 
identities 
\eqn\gamfour{t^{IJ}_{a\,b\,c\,d}\,t^{KL}_{a\,b\,c\,d}=
192\,\delta^{IL}\delta^{KJ}-48\,\delta^{IJ}\delta^{KL}+192\,\delta^{IK}\delta^{JL}\,,}
and for a more complicated case we have
\eqn\gamfive{\eqalign{&t^{MN}_{a\,b\,c\,d}\,\gamma^I_{a\,\dot e}\,
\gamma^J_{\dot e\,f}\,\gamma^K_{f\,\dot g}\,\gamma^L_{\dot g\,h}\,t^{PQ}_{h\,b\,c\,d}=\cr
&-96\,\delta^{ML}\delta^{PK}\delta^{NQ}\delta^{IJ}+96\,\delta^{MK}
\delta^{PL}\delta^{NQ}\delta^{IJ}
+96\,\delta^{MQ}\delta^{PL}\delta^{NK}\delta^{IJ}\cr&-96\,\delta^{MQ}
\delta^{PK}\delta^{NL}\delta^{IJ}
-96\,\delta^{ML}\delta^{PN}\delta^{QK}\delta^{IJ}-96\,\delta^{MP}
\delta^{NL}\delta^{QK}\delta^{IJ}\cr&
+96\,\delta^{MK}\delta^{PN}\delta^{QL}\delta^{IJ}+96\,\delta^{MP}
\delta^{NK}\delta^{QL}\delta^{IJ}
+96\,\delta^{ML}\delta^{PJ}\delta^{NQ}\delta^{IK}\cr&-96\,\delta^{MJ}
\delta^{PL}\delta^{NQ}\delta^{IK}
-96\,\delta^{MQ}\delta^{PL}\delta^{NJ}\delta^{IK}+96\,\delta^{MQ}
\delta^{PJ}\delta^{NL}\delta^{IK}\cr&
+96\,\delta^{ML}\delta^{PN}\delta^{QJ}\delta^{IK}+96\,\delta^{MP}
\delta^{NL}\delta^{QJ}\delta^{IK}
-96\,\delta^{MJ}\delta^{PN}\delta^{QL}\delta^{IK}\cr&-96\,\delta^{MP}
\delta^{NJ}\delta^{QL}\delta^{IK}
-96\,\delta^{MK}\delta^{PJ}\delta^{NQ}\delta^{IL}+96\,\delta^{MJ}
\delta^{PK}\delta^{NQ}\delta^{IL}\cr&
+96\,\delta^{MQ}\delta^{PK}\delta^{NJ}\delta^{IL}-96\,\delta^{MQ}
\delta^{PJ}\delta^{NK}\delta^{IL}
-96\,\delta^{MK}\delta^{PN}\delta^{QJ}\delta^{IL}\cr&-96\,\delta^{MP}
\delta^{NK}\delta^{QJ}\delta^{IL}
+96\,\delta^{MJ}\delta^{PN}\delta^{QK}\delta^{IL}+96\,\delta^{MP}
\delta^{NJ}\delta^{QK}\delta^{IL}\cr&
-96\,\delta^{ML}\delta^{PI}\delta^{NQ}\delta^{JK}+96\,\delta^{MI}
\delta^{PL}\delta^{NQ}\delta^{JK}
+96\,\delta^{MQ}\delta^{PL}\delta^{NI}\delta^{JK}\cr&-96\,\delta^{MQ}
\delta^{PI}\delta^{NL}\delta^{JK}
-96\,\delta^{ML}\delta^{PN}\delta^{QI}\delta^{JK}-96\,\delta^{MP}
\delta^{NL}\delta^{QI}\delta^{JK}\cr&
+96\,\delta^{MI}\delta^{PN}\delta^{QL}\delta^{JK}+96\,\delta^{MP}
\delta^{NI}\delta^{QL}\delta^{JK}
+192\,\delta^{MQ}\delta^{PN}\delta^{IL}\delta^{JK}\cr&-48\,\delta^{MN}
\delta^{PQ}\delta^{IL}\delta^{JK}
+192\,\delta^{MP}\delta^{NQ}\delta^{IL}\delta^{JK}+96\,\delta^{MK}
\delta^{PI}\delta^{NQ}\delta^{JL}\cr&
-96\,\delta^{MI}\delta^{PK}\delta^{NQ}\delta^{JL}-96\, \delta^{MQ}
\delta^{PK}\delta^{NI}\delta^{JL}
+96\,\delta^{MQ}\delta^{PI}\delta^{NK}\delta^{JL}\cr&+96\,\delta^{MK}
\delta^{PN} \delta^{QI}\delta^{JL}
+96\,\delta^{MP}\delta^{NK}\delta^{QI}\delta^{JL}-96\,\delta^{MI}
\delta^{PN}\delta^{QK}\delta^{JL}\cr&
-96\,\delta^{MP}\delta^{NI}\delta^{QK}\delta^{JL}-192\,\delta^{MQ}
\delta^{PN}\delta^{IK}\delta^{JL}
+48\,\delta^{MN}\delta^{PQ}\delta^{IK}\delta^{JL}\cr&-192\,\delta^{MP}
\delta^{NQ}\delta^{IK}\delta^{JL}
-96\,\delta^{MJ}\delta^{PI}\delta^{NQ}\delta^{KL}+96\,\delta^{MI}
\delta^{PJ}\delta^{NQ}\delta^{KL}\cr&
+96\,\delta^{MQ}\delta^{PJ}\delta^{NI}\delta^{KL}-96\,\delta^{MQ}
\delta^{PI}\delta^{NJ}\delta^{KL}
-96\,\delta^{MJ}\delta^{PN}\delta^{QI}\delta^{KL}\cr&-96\,\delta^{MP}
\delta^{NJ}\delta^{QI}\delta^{KL}
+96\,\delta^{MI}\delta^{PN}\delta^{QJ}\delta^{KL}+96\,\delta^{MP}
\delta^{NI}\delta^{QJ}\delta^{KL}\cr&
+192\,\delta^{MQ}\delta^{PN}\delta^{IJ}\delta^{KL}-48\,\delta^{MN}
\delta^{PQ}\delta^{IJ}\delta^{KL}
+192\,\delta^{MP}\delta^{NQ}\delta^{IJ}\delta^{KL}\,.}}
Although \gamfour-\gamfive\ were not used 
in this paper, they may prove to be very useful in future calculations. For a
more detailed list of properties of $SO(8)$ $\gamma$ matrices see \refs{\gsb}.
\vskip20pt
{\bf Acknowledgements:}
I am very grateful to M. Spradlin, L. Dolan and R. Rohm for useful discussions.
K.B. is partially supported by the U.S. Department of Energy,
Grant No. DE-FG02-97ER-41036/Task A.

\listrefs

\bye